\documentclass{report}

\usepackage{a4wide}
\usepackage{graphicx}

\def\nuc#1#2{\relax\ifmmode{}^{#1}{\protect\text{#2}}\else${}^{#1}$#2\fi}
\begin{document}

\begin{titlepage}

\newlength{\Size}
\setlength{\Size}{0.2\textwidth}
\newlength{\Shift}
\settoheight{\Shift}{L}
\addtolength{\Shift}{-\Size}

%%%%%%%%%%%%%%%%%%%%%%%%%%%%%%%%%%%%%%%%%%%%%%%%%%%%%%%%%%%%%%%%%%%%%%%%%%%%%%%%
%  the preprint number 
%%%%%%%%%%%%%%%%%%%%%%%%%%%%%%%%%%%%%%%%%%%%%%%%%%%%%%%%%%%%%%%%%%%%%%%%%%%%%%%%
\raisebox{\Shift}{\includegraphics{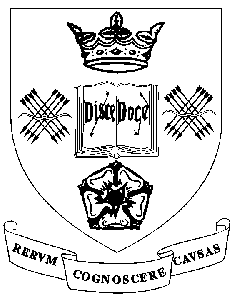}} 
\large\sf  
%%%%%%%%%%%%%%%%%%%%%%%%%%%%%%%%%%%%%%%%%%%%%%%%%%%%%%%%%%%%%%%%%%%%%%%%%%%%%%%%
%  the title 
%%%%%%%%%%%%%%%%%%%%%%%%%%%%%%%%%%%%%%%%%%%%%%%%%%%%%%%%%%%%%%%%%%%%%%%%%%%%%%%%
\vspace{2cm}
\begin{center}
\bf Neutron Shielding for Particle Astrophysics Experiments\bigskip \\
\end{center}
\normalsize\rm 
%%%%%%%%%%%%%%%%%%%%%%%%%%%%%%%%%%%%%%%%%%%%%%%%%%%%%%%%%%%%%%%%%%%%%%%%%%%%%%%%
%  the author list 
%%%%%%%%%%%%%%%%%%%%%%%%%%%%%%%%%%%%%%%%%%%%%%%%%%%%%%%%%%%%%%%%%%%%%%%%%%%%%%%%
\begin{center}
\rm J.E.McMillan\\ \em Department of Physics and Astronomy,
\\University of Sheffield\\
\rm j.e.mcmillan@sheffield.ac.uk 
\end{center}

\begin{center}
\rm \today    % <=== you may want to replace \today by a fixed date 
\end{center} 
%%%%%%%%%%%%%%%%%%%%%%%%%%%%%%%%%%%%%%%%%%%%%%%%%%%%%%%%%%%%%%%%%%%%%%%%%%%%%%%%
%  the abstract 
%%%%%%%%%%%%%%%%%%%%%%%%%%%%%%%%%%%%%%%%%%%%%%%%%%%%%%%%%%%%%%%%%%%%%%%%%%%%%%%%
\noindent 

{\bf Abstract:}
Particle astrophysics experiments often require large volume
neutron shields which are formed from hydrogenous material.
This note reviews some of the available materials in an attempt
to find the most cost effective solution. Raw polymer pellets
and Water Extended Polyester (WEP) shields are discussed in detail.
Suppliers for some materials are given.
\vspace*{\fill}
\begin{center}
\sf 
 
Department of Physics and Astronomy, 
University of Sheffield, 
Sheffield S3 7RH, UK
\end{center}
\end{titlepage}

%%%%%%%%%%%%%%%%%%%%%%%%%%%%%%%%%%%%%%%%%%%%%%%%%%%%%%%%%%%%%%%%%%%%%%%%%%%%%%%%
%  the text 
%%%%%%%%%%%%%%%%%%%%%%%%%%%%%%%%%%%%%%%%%%%%%%%%%%%%%%%%%%%%%%%%%%%%%%%%%%%%%%%%
\suppressfloats[t]   % no figure or table at the top of the page please 

\section*{Introduction}

Neutron shielding is required in current 
dark matter experiments both on detector systems and on neutron
source test and calibration systems.  In both applications, large
volumes are required, so low cost materials are desirable.
The ability to form the shielding material into arbitrary shapes, 
particularly voids in existing equipment, is important.

Detector systems operated underground must be shielded from environmental 
neutrons produced primarily by alpha-neutron knock-on reactions as a result 
of trace uranium and thorium in the rock surrounding the experimental 
caverns \cite{Ku02a}.
In this application, since neutron fluxes are low and since  
the detector systems discriminate against gammas, there is no requirement
to minimize (n-$\gamma$) production in the shield.  Low background 
materials are essential to ensure that the shield does not contribute to 
the detector background.  The ability to incorporate gadolinium, lithium
or other elements in a homogeneous manner might be required for future 
neutrino experiments. 

At the surface, experimental systems (and personnel) associated with neutron-beam or 
neutron-source tests  and calibrations of dark matter detectors require 
neutron shielding.  Here attenuation factors need to be as high as 
possible,
so it may be worth using boron
loaded material.  On some detector systems
it may be necessary to minimize (n-$\gamma$) production in the shield
so lithium
loading should also be considered as lithium
has a high neutron
absorption cross-section without gamma production.
Where 14MeV neutrons are produced, neutron 
activation of the shielding materials may become a problem.

The dimensions of shields depend primarily on the neutron attenuation 
factor required.   First order approximations can be obtained from the 
graphs and tables presented in the standard shielding texts 
\cite{Na71a,Ja68a,Ja75a,Sh00a}, while more detailed simulations are best 
performed with MCNP \cite{Br97a} as this has more detailed cross-section
data and better variance reduction capabilities than any of the alternative
montecarlo codes.

\section*{Hydrogenous Materials}
Fast neutrons, 1 to 15 MeV, are most effectively shielded by materials containing 
large amounts of hydrogen. The scattering of fast neutrons in collision with 
hydrogen atoms reduces the energies of the neutrons to epi-thermal 
and eventually to thermal energies where absorption cross-sections are 
much higher. 
The hydrogen may be held as water or 
as some hydrogenous organic compound.  
There are many hydrogenous materials which may be useful as neutron 
shielding.  Almost all these materials have a density of $\approx 1.0$,
\begin{table}
\centering
	\begin{tabular}{lll}
    water & &\pounds 0.00 /kg   \\
	paraffin oil  & &\pounds 0.76 /kg   \\
	paraffin wax  & &\pounds 1.15 /kg  \\
	polypropylene pellets  & Albis \cite{Al02a}&\pounds 0.48/kg  \\
    polyethylene slabs& John Caunt JC213 \cite{Ca02a} & \pounds 1.35/kg  \\
    polyethylene slabs& Barkston  \cite{Ba02a} & \pounds 2.11/kg  \\
    polyethylene slabs& IPSL \cite{Ip02a} & \pounds 2.53/kg \\
    borated polyethylene slabs& John Caunt JC201 \cite{Ca02a}& \pounds 2.13/kg \\
    boro-silicone  & John Caunt JC237 \cite{Ca02a}& \pounds 34.84/kg \\
	polypropylene slabs & IPSL \cite{Ip02a} & \pounds 3.46/kg  \\
	water extended polyester (WEP) & Scott Bader Crystic 1381PA \cite{Sc02a} &\pounds 1.02/kg\\
	WEP + polypropylene pellets &   &\pounds 0.70/kg\\
	resin neutron isolator & John Caunt JC243 \cite{Ca02a}& \pounds 47.54/kg \\
	neutron isolator & John Caunt JC244 \cite{Ca02a} & \pounds 7.92/kg \\
	\end{tabular}€
	\caption{Hydrogenous neutron shielding materials}
	\protect\label{tab1}
\end{table}€
Table \ref{tab1} shows a range of hydrogenous materials which have been 
considered for shielding, together with suppliers and current prices. 

Neutron-monitors, used for low energy cosmic ray modulation studies, were 
some of the earliest particle-astrophysics experiments to require neutron 
shielding.  A review of neutron-monitor designs is given by Hatton \cite{Ha71a} and 
it is interesting to note just how many of the considerations 
are still relevant to current experiments.  The primary 
requirement was to isolate the experiment from the local 
background neutron environment.   In the early designs 280mm thick 
paraffin wax shields were used, while in the later designs, extruded 
polyethylene slabs 75mm thick were used.   It was realized that at this 
thickness the experiment would not be completely isolated from its 
environment but the choice was a compromise based on the high cost of
the extruded material.   
The expense of extruded or cast slab polymer lies 
not only in the cost of the raw material, but also in the production of 
moulds or extrusion dies and the processing.  
It is still not easy to ensure that slabs are flat 
and void-free over the large scales required by typical experiments.

\section*{Polymer Pellets}

An alternative approach is to use raw polymer in small pellets,
as delivered by the polymer manufacturer, 
as a neutron shield \cite{Sh78a}.   Since the requirement is purely for hydrogen
solidified as polymer, there is no interest in melting point, 
average molecular weight or any of the other parameters used by the 
plastics industry. Even the ``high-density'' and
``low-density'' specifications used make no more 
than a few percent difference to the number of hydrogen atoms per unit 
volume.  Additionally, there appears little to choose between 
polyethylene and polypropylene pellets.
 This means that practically any material such as short runs, 
end-of-batch or similar, may be used to get the best possible price. 
Care should be taken that it is virgin pure polymer rather than pigmented or 
re-ground recycled material, primarily to avoid mineral fillers and the 
risk of contamination by trace radioactive material. 
Suitable polypropylene pellets have been obtained from 
Albis (UK) Ltd \cite{Al02a} at a price per kilo three or even four times 
less than the 
prices quoted for slab material.
Pelleted material has a measured packing fraction of about 0.6, so shields
made with this material must be made proportionately thicker than slab 
material.   The main disadvantage of pellets is that they are not 
load-bearing and some form of containment must be provided.

A system has been developed for NAIAD in which the lead and copper 
shielding castle is surrounded by an 150mm thick jacket of polypropylene 
pellets.  These pellets are held in woven polypropylene fabric sacks 
800mm$\times$100mm which were filled in-situ. On the outside of the neutron shield 
is a retaining wall of plywood held in place by steel structural frame.  
The sacks full of pellets are located between the lead and the retaining 
wall and the spaces between sacks filled with additional pellets.  The 
pellets are simple to position by pouring and can easily be removed with 
an industrial vacuum cleaner.  This modular construction technique was 
found simple to assemble and offers the possibility of easy disassembly 
and re-use of the material on future experiments.

\section*{Water Extended Polyester}
Oliver and Moore \cite{Ol70a} recognized that water extended polyester 
(WEP) was a  promising material for neutron shielding since it was 
essentially 50\% water held in a solid form.
This material is a liquid thermosetting polyester resin into which water 
can be incorporated to form a thick emulsion which then hardens to a 
material similar to a fine-grained plaster. 
The droplets of the emulsion are in the range 1--$5\mu$m and when cured, 
the aqueous phase remains trapped in these droplets within the rigid
polyester matrix.
The final material can be easily drilled or machined and will bear substantial loads.
A substantial shield for a $^{252}$Cf source fabricated from WEP is 
described by Veerling \cite{Ve73a}.
Two currently available commercial versions of WEP have been identified, 
namely Aropol TM WEP 662P
from Ashland Chemical \cite{As02a} and Crystic 1381PA from Scott-Bader 
\cite{Sc02a}.

\subsection*{Crystic 1381PA}
Experiments have been performed using 
Crystic 1381PA, a material which will blend with more than 60\% water and still harden.
The polyester 
contains carbon, oxygen and carbon, and while the manufacturer's notes give no
details, other polyesters are in the ratio C$_{5}$H$_{4}$O$_{2}$, while the 
material used in \cite{Ol70a} is quoted as C 25.3\% H 9.7\% and O 65\% by 
weight when cured.  The 
resin also contains an emulsifying agent, 
\cite{Ol70a} mentions
cobalt octoate and dimethylanaline, so traces of both cobalt and nitrogen may be expected.  
The curing catalyst is primarily methylethyl ketone peroxide, a 
powerful and extremely unpleasant oxidising agent.

The uncured material contains a high proportion of styrene monomer
which has a pungent aromatic odour and consequently
the mixing and preparation of WEP 
must be done in a well ventilated building, or preferably outdoors during 
fine weather.   The styrene is fully polymerized during curing and the 
resulting 
material has no residual odour.  
During curing, considerable heat is 
evolved and consequently it cannot be cast in thicknesses greater than 
about 100mm.  Exceeding this will cause loss of water 
and in extreme cases, the polyester will begin to degrade and even char.

The resin,
as supplied, must be blended with  water using a high-shear blender.  
On 
an experimental scale, any domestic blender capable of making mayonnaise 
can be used, on a medium scale an industrial blender would work, while a DOE 
study \cite{Do99a} describes systems capable of producing more than one 
tonne per day. 
When the water has been satisfactorily blended, the curing catalyst 
is added and thoroughly dispersed in the material.   The resin emulsion can
then be poured into moulds and allowed to set. This happens in about 15 
minutes if the specified ratios are used.   Cleaning the blender 
and containers is not difficult but must be done well before the material 
begins to harden.  The costs quoted in table \ref{tab1} are the raw material 
costs without any 
consideration of the expenses of labour or mould production.  

An experiment was performed in which Crystic 1381PA was mixed with polypropylene 
pellets, in order to avoid the packing fraction problem encountered when 
only pellets are used.  
The result was a solid concrete-like composite
material in which the aggregate was polypropylene pellets. While little 
care was taken in the production, the result contained very few entrapped 
air bubbles.   This composite is clearly considerably cheaper to produce 
than pure WEP. It would make an attractive material for use where sacks of 
pellets are not feasible or where a low cost load bearing shielding 
material is required.

No experiments have yet been performed on the shielding properties of 
Crystic 1381PA, though the original work of Oliver and Moore \cite{Ol70a}
gives some measurements.  These and considerations of the constituents
of WEP suggest that its neutron shielding capabilities will be similar 
to those of PMMA(Lucite, Perspex).
\begin{table}
\centering
	\begin{tabular}{lll}
	Colemanite, hydrated calcium borate hydroxide CaB$_{3}$O$_{4}$(OH)$_{3}\cdot$H$_{2}$O
 & Bainbridge\cite{Ba02b} & \pounds 1.07/kg \\
	Lithium carbonate, presumably fairly pure (flux grade) Li$_{2}$CO$_{3}$& Bainbridge\cite{Ba02b} & \pounds 9.08/kg \\
	Petalite, lithium aluminium silicate Li$_{2}$OAl$_{2}$O$\cdot$SiO$_{2}$& Bainbridge\cite{Ba02b} & \pounds 0.96/kg \\
	Borax (domestic), hydrated sodium borate Na$_{2}$B$_{4}$O$_{7}\cdot$10H$_{2}$O & Boots & \pounds 1.63/kg \\
	\end{tabular}
	\caption{Inorganic neutron capture materials}
	\protect\label{tab2}
\end{table}

WEP can easily be blended with inorganic salts
or oxides to provide loaded hydrogenous shielding.   These inorganic 
materials can either be inert powders or dissolved in the water phase.
It is not thought that inorganics will affect the hardening or emulsifying
properties of the resin, provided that strong reducing agents are avoided.
Oliver and Moore \cite{Ol70a} performed experiments with dissolved sodium 
metaborate and boric acid which had no effect on hardening.  They also 
claim that the emulsion was sufficiently stiff to withstand the 
incorporation of lead shavings without settling.  Recent work commissioned by the 
DOE \cite{Do99a}, in which WEP was successfully used to 
encapsulate radioactively contaminated sludges and salt solutions, indicates that inorganic
materials can be blended and that the expected lifetimes of the resulting 
resins are high.

For neutron shielding, the materials most used would be compounds of 
Boron and Lithium, both of which have large capture cross-sections at 
thermal energies.  Clearly, for shielding applications, there is little 
requirement for chemical purity, so domestic materials or finely powdered 
minerals intended for ceramic glazes may be used with considerable 
savings in cost.  Table~\ref{tab2} shows some materials which have been 
identified as potential loading materials for WEP neutron shields.
 
\section*{Disclaimer}
The suppliers mentioned are primarily British while the prices quoted are 
those of 2002/3.  The lists are in no sense exhaustive. 
  
\nocite{*}
\bibliographystyle{plain}
%\addcontentsline{toc}{section}{Bibliography}
\begin{flushleft}
\bibliography{shielding}
\end{flushleft}
\end{document}